\documentclass[twocolumn,showpacs,preprintnumbers,amsmath,amssymb]{revtex4}
\usepackage{amsmath}
\usepackage{graphicx}
\begin{document}

\title{\bf Phenomenological theory of spin excitations in La- and Y-based cuprates}

\author
{
 T. Zhou$^{1}$ and Z. D. Wang$^{1,2}$
}

\affiliation{$^{1}$Department of Physics, and Center of Theoretical
and Computational Physics, The University of Hong Kong, Pokfulam
Road, Hong Kong, China\\
$^{2}$National Laboratory of Solid State Microstructures, Nanjing University, Nanjing 210093, China\\
}

\date{\today}

\begin{abstract}
Motivated by recent inelastic neutron scattering (INS) experiments
on La-based cuprates and based on the fermiology theories, we study
the spin susceptibility for La-based (e.g., La$_{2-x}$Sr$_x$CuO$_4$)
and Y-based (e.g., YBa$_2$Cu$_3$O$_y$) cuprates, respectively. The
spin excitation in YBa$_2$Cu$_3$O$_y$ is dominated by a sharp
resonance peak at the frequency 40 meV in the superconducting state.
Below and above the resonance frequency, the incommensurate (IC)
peaks develop and the intensity of the peaks decreases dramatically.
In the normal state, the resonant excitation does not occur and the
IC peaks are merged into commensurate ones. The spin excitation of
La$_{2-x}$Sr$_x$CuO$_4$ is significantly different from that of
Y-based ones, namely, the resonance peak does not exist due to the
decreasing of the superconducting gap and the presence of the
possible spin-stripe order. The spectra are only enhanced at the
expected resonance frequency (about 18 meV) while it is still
incommensurate. On the other hand, another frequency scale at the
frequency 55 meV is also revealed, namely the spectra are
commensurate and local maximum at this frequency. We elaborate all
the results based on the Fermi surface topology and the $d$-wave
superconductivity, and suggest that the spin-stripe order be also
important in determining the spin excitation of La-based cuprates. A
coherent picture for the spin excitations is presented for Y-based
and La-based cuprates.

\end{abstract}
\pacs{74.25.Ha, 74.20.Mn}

 \maketitle

\section{introduction}

Spin excitations in high-$T_c$ superconductors have been intensively
studied by the inelastic neutron scattering (INS) experiments in the
past decade. Most of the INS experiments are performed on
YBa$_2$Cu$_3$O$_y$ (YBCO) and La$_{2-x}$Sr$_x$CuO$_4$ (LSCO)
samples. One of the most important results revealed by the
experiments in YBCO is the resonant spin
excitation~\cite{ros,fon,moo,fon1,dai,sto}, which is centered at the
momentum $(\pi,\pi)$ and the intensity decreases dramatically as the
momentum deviates from $(\pi,\pi)$. As the frequencies are below and
above the resonance frequency, the spin excitations are
incommensurate (IC) and the peaks disperse towards the momentum
$(\pi,\pi)$ as the frequencies are close to the resonance
frequency~\cite{hay,ara,bou}. Above the superconducting transition
temperature or possibly above the pseudogap temperature, the
resonant excitation disappears and the spectrum  are commensurate at
all the frequencies~\cite{ara,bou,dai1}. Besides YBCO, the resonance
peak is also observed in
Bi$_2$Sr$_2$CaCu$_2$O$_{8+x}$~\cite{fong,roso},
Tl$_2$Ba$_2$CuO$_{6+x}$~\cite{he}, and electron-doped samples
Pr$_{0.88}$LaCe$_{0.12}$CuO$_{4-x}$~\cite{step}, with the resonance
frequencies scaling approximately with superconducting transition
temperature $T_c$ in different systems~\cite{step,roso} as well as
in the same system with different doping
densities~\cite{moo,fon1,dai}.

Recent experiments revealed that the dispersion of the IC peaks in
La-based ones is similar to that of the Y-based ones, namely the
spin excitation is commensurate at the frequency $\omega_c\sim 50$
meV, where a downward dispersion and an upward dispersion are seen
below and above $\omega_c$, respectively~\cite{tra,chr}. On the
other hand, the spin excitation of La-based ones is significantly
different from that of Y-based ones. Firstly, the peak approaches to
$(\pi,\pi)$ only when the frequency is close to $\omega_c$ and the
dispersion is weakly dependent on the frequency at low
frequencies~\cite{tra,chr}. Secondly, the IC spin excitation is
observed even when the temperature is well above the superconducting
transition temperature~\cite{thu,mas,aep,fuj}. More importantly, the
spin resonance peak, which is expected to occur at  the frequency
around $18$ meV (scaled with $T_c$) in optimal doped LSCO, has not
been observed experimentally. The intensity of the commensurate peak
at the frequency $\omega_c$ is not strongest compared to the IC
ones, which is also different from that of the spin resonance peak
in Y-based systems. As the temperature crosses the transition
temperature $T_c$, the intensity is not enhanced suddenly
either~\cite{tra,chr}. Thus the commensurate spin excitation is also
significantly different from the resonant excitation. In addition,
very recently, the experiment on LSCO sample revealed two frequency
scales in the superconducting state~\cite{vig}, i.e., at low
frequencies, the intensity of the IC peak increases as the frequency
increases and reaches the maximum value at the frequency around $18$
meV; while the spin excitation exhibits a broad hump with the
commensurate spin excitation occurring at $50$ meV and the intensity
reaches the local maximum value at this frequency. Interestingly,
the low frequency scale is just the expected resonance frequency in
LSCO, which also suggests that there should be a coherent picture
for the spin excitations of YBCO and LSCO.

Theoretically, there have been two possible explanations for the
spin excitations observed in the INS experiments. One is based on
the fermiology
theories~\cite{si,tana,tana1,liu,bri,morr,kao1,norman,jxli,li,jxl,zhou,ere,kao,schn,hir,dahm,mans,zhou1},
namely, the resonance peak is a collective spin excitation mode, and
the IC excitation is caused by the nested Fermi surface. This
scenario is rather popular in describing spin excitations of YBCO
materials. For La-based samples, it was proposed that the
disappearance of the resonance is due to the different shape of the
Fermi surface, namely, the Fermi surface is suggested to be centered
at $(0,0)$ rather than $(\pi,\pi)$ for La-based cases, which is
caused by the decrease of the nearest neighbor hoping constant
$t^{\prime}$~\cite{si,tana,tana1,liu}. However, angel resolved
photoemission spectroscopy (ARPES) experiments have shown that the
Fermi surface of LSCO looks similar to that of Y-based ones, i.e.,
it is also centered at $(\pi,\pi)$, and the nearest neighbor hoping
constant $t^{\prime}$ in LSCO samples is about $0.25\sim0.3$ t,
which does not cause a qualitative change of the shape of the Fermi
surface~\cite{ino,ino1,yos}. To be consistent with the resent ARPES
experiments, a picture based on a distorted Fermi surface, i.e., the
Fermi surface expands along the $k_x$ axis and shrinks along the
$k_y$ axis, is proposed to explain some of the characteristic of the
magnetic excitation of LSCO~\cite{yam}. Based on this scenario and
the random phase approximation (RPA), the spin resonance is possibly
very weak, depending on the choice of the RPA factor. While the
quasi-resonance peak at $(\pi,\pi)$ still exists even if the RPA
renormalized factor $r$ [Ref.~\cite{yam}] is much smaller than
Y-based ones. So, particularly for La-based samples, an explanation
that suggests the presence of the dynamic stripes with a kind of
4-lattice constant charge order and 8-lattice constant spin order to
be the origin of the IC peaks, has attracted much
interest~\cite{kru,car,voj,uhr,uhr1,and,sei,sei1,yao}.  This picture
offers a natural explanation for the observation of a charge order
signal whose wave vector is just twice of the IC magnetic wave
vector~\cite{tra1,ich,fuj1}.

The IC spin fluctuations associated with a stripe phase are expected
to be one dimensional, i.e., the IC peaks appear either at $(q,\pi)$
or $(\pi,q)$ direction, while most INS experiments are in fact
performed on twinned samples, thus along a given $a$ or $b$
directions, the domains with lattice spacing $a$ or $b$ exist in
equal proportion. So, all asymmetries between the $a$ and $b$
direction will be covered up. The $a-b$ anisotropic IC spin
excitation was first reported by Mook $et$ $al$. in partly detwinned
YBa$_2$Cu$_3$O$_{6.6}$ and was seen as a strong support for the
stripe phase picture~\cite{mook}. While later INS experiments on
fully untwinned YBCO samples revealed that the IC peaks are actually
two dimensional although a clear $a-b$ anisotropy exists~\cite{hin}.
Thus an alternative explanation for the $a-b$ anisotropy in YBCO
samples is based on the nested fermi surface scenario, i.e., by
taking into account the role of the CuO chain or distorted fermi
surface~\cite{zhou,ere,kao,schn,hir}. The $a-b$ anisotropic IC spin
excitation was also reported in La-based systems recently. This
anisotropic feature is remarkably different from that of the YBCO
and is actually one-dimensional, so that it supports strongly the
presence of the spin-stripe order in this system~\cite{chri}.

In this paper, motivated by the above observations on La-based
samples, we study the spin excitation of Y-based and La-based
cuprates based on the fermiology theories. We attempt to understand
the features of the spin excitation in LSCO by comparing
 the similarities and differences of magnetic excitations
between YBCO and LSCO. Apart from the topology of the Fermi surface,
the spin-stripe order seems to play an important role in the spin
excitation of La-based cuprates. Thus for La-based cases, we
phenomenologically take into account a possible 1/8-lattice spin
order by using an IC wave vector in the vertex of the RPA factor.
Since this picture can also explain the spin dynamics of the bilayer
samples~\cite{zhou1} and the electron-doped ones~\cite{jxl}, we here
give a coherent picture of the spin dynamics of the high-$T_c$
superconductors.

The article is organized as follows. In Sec. II, we introduce the
model and work out the formalism. In Sec. III, we present numerical
results of the spin susceptibility for La-based and Y-based cases,
respectively. In Sec. IV, we interpret the results. Finally, we give
a brief summary in Sec. V.

\section{Model and formalism}
We start with a BCS bare spin susceptibility in a one-layer
superconducting system,

\begin{eqnarray}
\chi_0({\bf q},\omega) &=& \frac{1}{N}\sum_{\bf
k}\{\frac{1}{2}[1+\frac{\varepsilon_{\bf k}\varepsilon_{{\bf k}+{\bf
q}}+\Delta_{\bf k}\Delta_{{\bf k}+{\bf q}}}{E_{\bf k}E_{{\bf k}+{\bf
q}}}]\nonumber\\&&\times \frac{f(E_{{\bf k}+{\bf q}})-f(E_{\bf
k})}{\omega-(E_{{\bf k}+{\bf q}}-E_{\bf k})+i\Gamma}
+\frac{1}{4}\nonumber\\&& \times [1-\frac{\varepsilon_{\bf
k}\varepsilon_{{\bf k}+{\bf q}}+\Delta_{\bf k}\Delta_{{\bf k}+{\bf
q}}}{E_{\bf k}E_{{\bf k}+{\bf q}}}] \nonumber\\&& \times
\frac{1-f(E_{{\bf k}+{\bf q}})-f(E_{\bf k})}{\omega+(E_{{\bf k}+{\bf
q}}+E_{\bf k})+i\Gamma} +
\frac{1}{4}\nonumber\\&&\times[1-\frac{\varepsilon_{\bf
k}\varepsilon_{{\bf k}+{\bf q}}+\Delta_{\bf k}\Delta_{{\bf k}+{\bf
q}}}{E_{\bf k}E_{{\bf k}+{\bf q}}}]\nonumber\\&& \times
\frac{f(E_{{\bf k}+{\bf q}})+f(E_{\bf k})-1}{\omega-(E_{{\bf k}+{\bf
q}}+E_{\bf k})+i\Gamma} \},
\end{eqnarray}
where $f(E_{\bf k})$ is the Fermi distribution function, $E_{\bf
k}=(\varepsilon_{\bf k}^{2}+\Delta_{\bf k}^{2})^{1/2}$ with
$\varepsilon_{\bf k}$ and $\Delta_{\bf k}$ the electron band
dispersion and superconducting gap function, respectively.

For different materials, different maximum gaps ($\Delta_0$) are
chosen, i.e., at the optimal doping, the superconducting gap
magnitude ($\Delta_{max}$) of the different systems at the zero
temperature limit scales linearly with the corresponding transition
temperature $T_c$~\cite{sato,sato1,feng}, expressed by
$2\Delta_{max}=5.5$ $T_c$. The bare normal state spin susceptibility
is obtained by setting $T=T_c$ and $\Delta_{\bf k}\equiv 0$. The
band dispersion is used by fitting qualitatively the Fermi surface
as measured by ARPES~\cite{norm}, written as $\varepsilon_{\bf
k}=\sum t_i \eta_i$, with $t_{0-5}=130.5$, $-595.1$, $163.6$,
$-51.9$, $-111.7$, $51$ (meV), and $\eta_{0-5}=1$, $(\cos k_x+\cos
k_y)/2$, $\cos k_x \cos k_y$, $(\cos 2k_x+\cos 2k_y)/2$, $(\cos k_x
\cos 2k_y+\cos 2k_x\cos k_y)/2$, $\cos 2k_x \cos 2k_y$. The
superconducting gap function is taken as $d$-wave symmetry with
$\Delta_{\bf k}=\Delta_0(\cos k_x-\cos k_y)/2$ and $\Delta_0$ being
obtained by the corresponding $T_c$.

Taking into account the electron-electron interaction, the
renormalized spin susceptibility is given by a RPA form,
\begin{equation}
\chi({\bf q},\omega)=\frac{\chi_0({\bf q},\omega)}{1-U_{\bf \mathcal
{Q}}({\bf q}) \chi_0({\bf q},\omega)},
\end{equation}
with $U_{\bf \mathcal {Q}}({\bf q})$ is the spin-spin response function and expressed by,
\begin{equation}
U_{\bf \mathcal {Q}}({\bf q})=U[\cos(q_x + \mathcal {Q}_x)+\cos(q_y
+ \mathcal {Q}_y)]/2,
\end{equation}
where $U$ is fixed at 210 meV in the following calculations. The
wave vector ${\bf \mathcal {Q}}$ represents the effect of the spin
order. For $t-J$ type model and probably suitable for most families
of cuprates, ${\bf \mathcal {Q}}={\bf Q}=(\pi,\pi)$, thus $U_{\bf
\mathcal {Q}}=-U(\cos q_x+\cos q_y)/2$, which tends to suppress the
incommensurability since it is largest at the commensurate wave
vector $(\pi,\pi)$. While for the La-based samples, we use ${\bf
{\mathcal {Q}}}={\bf Q_1}=(\pi\pm 2\pi/8,\pi)$, by taking into
account the possible 1/8-spin order ($\pm$ depends on whether $q_x$
is greater than $\pi$ to ensure the spectra is symmetric along the
line $q_x=\pi$), so that $U_{\bf \mathcal {Q}}$ is largest at the IC
wave vector ${\bf Q_1}$.

\section{results}
\subsection{Evolution of the spin resonance for different systems}

\begin{figure}
\centering
\includegraphics[scale=0.5]{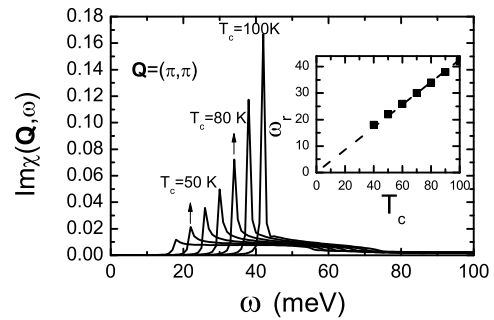}
\caption{The imaginary part of the spin susceptibility as a function
of the frequency in the superconducting state, with $T_c$ varying
from 40 K to 100 K. The inset is the resonance frequency vs. $T_c$.
 }\label{fig1}
\end{figure}

The renormalized spin susceptibility  as a function of the frequency
for different superconducting transition temperatures ($T_c$) in the
superconducting state with $\mathcal {Q}=(\pi,\pi)$ is plotted in
Fig. 1. As is seen, the resonance frequency $\omega_r$ decreases as
$T_c$ decreases. Moreover, as seen in the inset of Fig. 1,
$\omega_r$ is in fact proportional to $T_c$, i.e., $\omega_r\approx
5 T_c$, qualitatively consistent with the experimental
results~\cite{step}. As reported by earlier
experiments~\cite{fon1,roso,dai} and theoretical
calculations~\cite{dahm,mans,zhou1}, this linear relation also holds
as the level doping decreases from the optimal doping. Thus in
cuprates, the relation for the resonance frequency $\omega_r$ and
$T_c$ is in fact independent on the systems or doping densities. On
the other hand, we can see clearly from Fig. 1 that the intensity of
the resonance peak decreases as $T_c$ decreases. Note that, for the
samples with $T_c = 40$ K, only quasi-resonance occurs with a much
weaker intensity. Therefore the resonance signals are not so clear
for LSCO as for YBCO even if the stripe order and the band
difference are neglected. As we will see below, the resonance in
LSCO is suppressed further by the 1/8-spin order and becomes
incommensurate at the expected resonance or quasi-resonance
frequency.

\subsection{Spin susceptibility of LSCO}
Now let us look into the spin susceptibility of LSCO with $T_c=40$ K
and $\mathcal {Q}={\bf Q_1}$. The imaginary parts of the spin
susceptibilities at the wave vector ${\bf Q}=(\pi,\pi)$ and IC wave
vector ${\bf Q^{\prime}}=(0.85\pi,\pi)$ in the normal and
superconducting states are plotted in Fig. 2, respectively. As seen,
the spectra show a peak-dip-hump structure with two local maximums
at about 18 meV and 55 meV in the superconducting state. The spin
excitation at the frequency 55 meV may be commensurate because the
intensity decreases as the momentum shifts from ${\bf Q}$ to ${\bf
Q^{\prime}}$. The peak intensity at this frequency is slightly
weaker than that of the IC one at 18 meV. This two component feature
is well consistent with a very recent experiment~\cite{vig}. In the
normal state, the intensity at the frequency 18 meV decreases, and
only one maximum at the frequency 50 meV exists, while many features
are still similar to that of the superconducting state, namely the
spin excitation is incommensurate at low and high frequency and the
incommensurability is very small at the frequency 50 meV.

\begin{figure}
\centering
\includegraphics[scale=0.5]{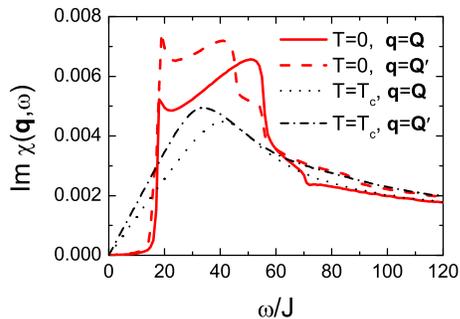}
\caption{(Color online) The imaginary part of the spin
susceptibility as a function of the frequency at the commensurate
momentum ${\bf Q}$ and IC momentum ${\bf Q^{\prime}}$ in the
superconducting and normal states, respectively.
 }\label{fig2}
\end{figure}

The intensity plots of the imaginary parts of the spin
susceptibilities as functions of the momentum and frequency in the
superconducting and normal states are plotted in Figs. 3(a-d),
respectively. At low frequencies, the spin excitation is along the
diagonal direction. A clear spin gap exists along the parallel
direction [Fig. 3(a)]. When the frequency increases, the spin
excitation along parallel direction is available and dominates over
that along diagonal direction for the frequency above 10 meV. As the
frequency reaches about 18 meV, the IC spin excitation reaches the
maximum intensity. Then the intensity decreases as the frequency
increases. Note that the incommensurability depends weakly on the
frequency when the frequency is less than 40 meV. As the frequency
increases further the intensity of the peak increases again and the
peak position approaches to $(\pi,\pi)$ quickly. At the frequency 55
meV, the spin excitation is commensurate and the intensity reaches
the local maximum. As the frequency is above 55 meV, the IC peaks
reappear and are dispersing with a upward curvature. The dispersion
does not change much and  has also a hourglass shape in the normal
state, as seen in Figs. 3(c-d). Meanwhile the spin gap is absent in
the normal state. These results agree well with experimental results
in La-based ones~\cite{thu,mas,aep,fuj,tra,chr}.

\begin{figure}
\centering
\includegraphics[scale=0.45]{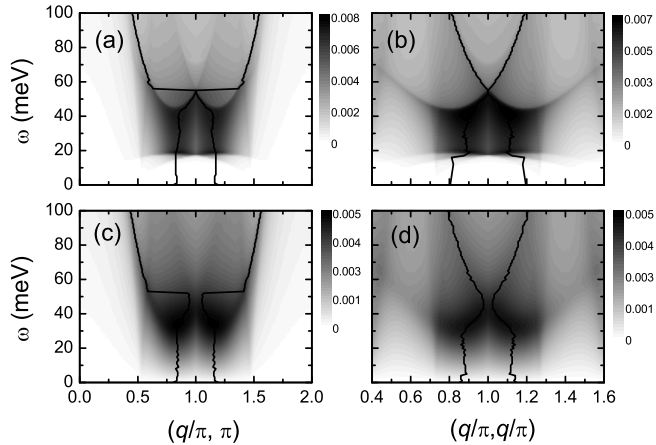}
\caption{The intensity plots for imaginary parts of the spin
susceptibilities as functions of the momentum and frequency in the
superconducting (a-b) and normal (c-d) states, respectively.
 }\label{fig3}
\end{figure}

We turn to address a possible anisotropy caused by the 1/8-spin
order. The intensity plot of the imaginary part of the spin
susceptibility as a function of the momentum is shown in Fig. 4. The
one dimensional IC peaks along $q_x$ direction are seen clearly, and
the spin excitation is commensurate along $q_y$ direction. The
direction of the IC excitation is determined by the wave vector
$\mathcal {Q}$. This $a-b$ anisotropy caused by the spin order is
significantly different from that observed in untwinned YBCO
samples~\cite{hin}, namely in untwinned YBCO samples, the spin
excitation is actually two dimensional although the IC peak along
$(q,\pi)$ direction is stronger. In La-based cuprates, such
one-dimensional spin excitation spectra are reported
recently~\cite{chri}. Interestingly, the present phenomenological
picture can reproduce some of the experimental results, particularly
the one-dimensional spin excitation. In fact, the observation of the
one-dimensional IC spin excitation supports strongly the presence of
the stripe order and suggest that it should play an important role
in determining the spin excitation spectra in La-based systems.

\begin{figure}
\centering
\includegraphics[scale=0.5]{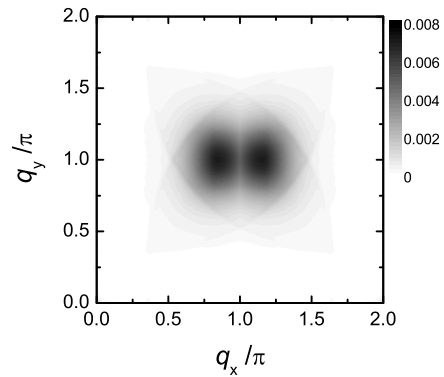}
\caption{The intensity plot for the imaginary part of the spin
susceptibility as a function of the momentum with the frequency
$\omega=20$ meV.
 }\label{fig4}
\end{figure}

\subsection{Spin susceptibility of YBCO}

For the case of YBCO samples with $T_c=90$ K and $\mathcal {Q}={\bf
Q}$, the imaginary parts of the spin susceptibilities Im$\chi$ as a
function of the frequency in the superconducting and normal states
are plotted in Fig. 5, respectively. In the superconducting state,
Im$\chi$ is dominated by a sharp resonance at the frequency 40 meV.
In the normal state, the intensity decreases dramatically while it
still has a broad peak at the frequency 30 meV. This result is
significantly different from that of La-based ones. Besides, there
is only one frequency scale in the superconducting state in YBCO,
i.e., the resonance frequency.
Below and above the resonance frequency the intensity decreases
dramatically.

\begin{figure}[htb]
\centering
\includegraphics[scale=0.5]{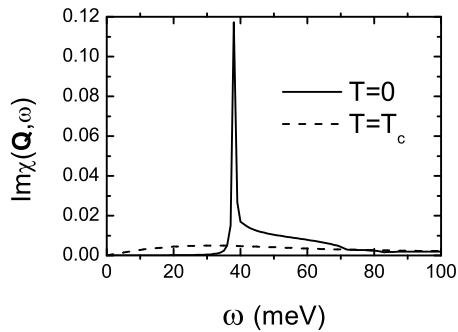}
\caption{The same as that of Fig.2 at the momentum $(\pi,\pi)$, but
with $T_c=90$ K and ${\bf \mathcal {Q}}={\bf Q}$.
 }\label{fig5}
\end{figure}
\begin{figure}
\centering
\includegraphics[scale=0.45]{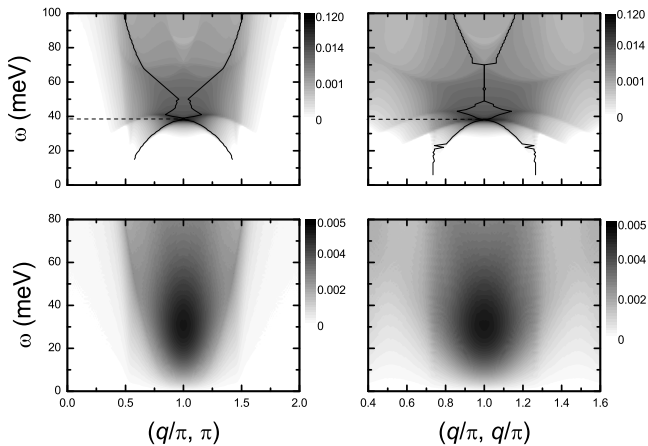}
\caption{The same as that of Fig.3, but with $T_c=90$ K and ${\bf
\mathcal {Q}}={\bf Q}$.
 }\label{fig6}
\end{figure}

The intensity plots for imaginary parts of the spin susceptibilities
as functions of the momentum and frequency in the superconducting
and normal states are shown in Figs. 6(a-d), respectively. As seen,
the spin susceptibility is peaked at the momentum $(\pi,\pi)$ at 40
meV in the superconducting state. Below the resonance frequency,
dominant IC peaks at the momentum $(\pi\pm\delta,\pi)$ and a
subdominant structure along the diagonal lines occur. As the
frequency is low enough, the IC peaks rotate to the diagonal
direction due to the spin gap along the parallel direction. The
peaks are dispersing with a downward curvature. Above the resonance
frequency the IC peaks reappear and disperse with an upward
curvature. Above the frequency 70 meV, the IC peaks along the
diagonal direction occur and dominate over the IC peaks along the
parallel direction. This hourglass dispersion is similar to that of
the LSCO. While the incommensurability depends strongly on the
frequency, different from that of the La-based cases, mainly due to
the the different wave vector ${\bf \mathcal {Q}}$. The maximum
intensity of the spin susceptibility is around the frequency 30-40
meV in the normal state. A pronounced distinction between the YBCO's
and LSCO's spectra in the normal state is the peak dispersion,
namely, for YBCO's cases, the spin susceptibility is commensurate at
all the frequencies considered.

\section{discussions}

\begin{figure}
\centering
\includegraphics[scale=0.52]{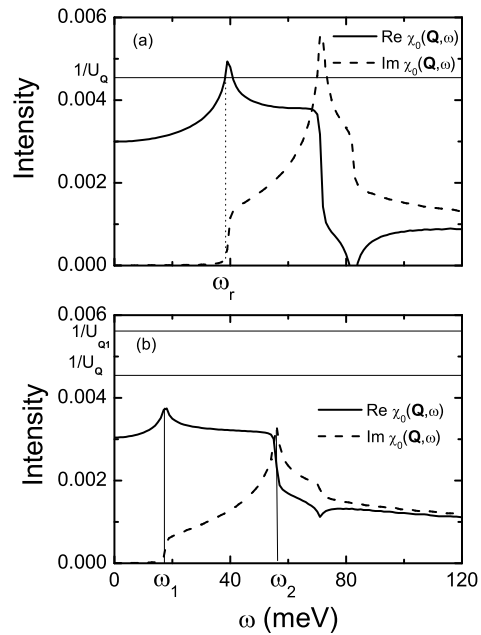}
\caption{The imaginary and real parts of the bare spin
susceptibility as a function of the frequency in the superconducting
state at the momentum $(\pi,\pi)$ for $T_c=90$ K and $40$ K,
respectively.
 }\label{fig7}
\end{figure}

At this stage, we attempt to elaborate the origin of the above
features based on the topology of the Fermi surface. The
renormalized spin susceptibility [Eq.(2)] consists of two
components, i.e., the bare spin susceptibility $\chi_0$ and the RPA
factor [$1+U_{\mathcal {Q}}\chi_0$]. The imaginary and real parts of
the bare spin susceptibilities for $T_c=90$ K and $40$ K are plotted
in Fig. 7(a) and 7(b), respectively. At low frequencies, the
imaginary part of the bare spin susceptibility at the momentum
$(\pi,\pi)$ approaches to zero due to the presence of the
superconducting gap, as seen in Fig. 7. Thus the renormalized spin
susceptibility also approaches to zero. It has a step-like rise as
the frequency approaches to $2\Delta_0$ due to the flat band near
$(\pi,0)$. As a result, the real part of the bare spin
susceptibility Re$\chi_0$ develops a sharp structure and reaches the
maximum at this frequency. For the samples with $T_c=90$ K, a pole
occurs, namely, the real part of the RPA factor $1+U_{{\bf \mathcal
{Q}} }$Re$\chi_{0}$ equals to zero at the frequency $\omega_r$, as
seen in Fig. 7(a). In fact, the RPA factor plays a major role and
the imaginary part of the renormalized spin susceptibility equals to
Re$\chi_0$/Im$\chi_0$ at this frequency. As the imaginary part of
the spin susceptibility is small due to the spin gap, this suggests
the formation of a spin collective mode, ascribed to be the spin
resonance. The resonance peak is very sharp so that the peak
intensity decreases dramatically as the frequency increases, as seen
in Fig. 5. For the case of $T_c=40$ K, the maximum of the real part
of the bare spin susceptibility is much smaller than that of the
$T_c=90$ K's sample due to the decrease of the superconducting gap.
So the pole condition cannot be satisfied even if the stripe order
is absent. Furthermore, $1/U_{\mathcal {Q}}({\bf Q})$ decreases as
$\mathcal {Q}$ becomes ${\bf Q_1}$, as seen in Fig. 7(b). Thus the
existence of the spin-stripe order suppresses further the peak
intensity at the expected resonance frequency and the IC peaks are
developed at this frequency. While the peak intensity at the
frequency $\omega_1$ is still enhanced by the RPA factor. On the
other hand, we can see from Fig. 7(b) that the imaginary part of the
bare spin susceptibility has a maximum intensity at the frequency
$\omega_2$. Because the peak intensity of the renormalized spin
susceptibility at the frequency $\omega_1$ is not as strong and
sharp as that of the resonance peak in YBCO samples, the higher
frequency component caused by the bare spin susceptibility can still
be seen (Fig. 2 and Fig. 3). Thus the spin excitation of LSCO shows
a clear feature of two frequency components, with the maximum
intensity being at $\omega_1$ following a broad peak at the
frequency $\omega_2$. In fact, it is the competition between the
bare spin susceptibility and the RPA factor that determines the
feature of the renormalized spin susceptibility in LSCO. The role of
RPA becomes smaller and smaller as $T_c$ decreases, and is
suppressed further by the spin-stripe order. For LBCO samples, the
transition temperature is merely $20$ K so that the maximum of the
real part of the spin susceptibility is even smaller. Meanwhile at
some doping the superconductivity is suppressed completely, and thus
a strong stripe order may emerge in this sample~\cite{mood,vall};
therefore the RPA factor is not important even near the frequency
$2\Delta_0$, such that the lower frequency compound is suppressed
and only the higher frequency component with the commensurate spin
excitation at the frequency 50 meV is observed~\cite{tra}.

\begin{figure}
\centering
\includegraphics[scale=0.5]{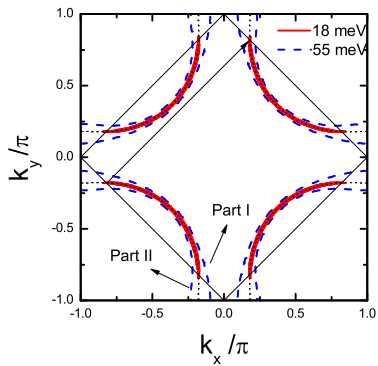}
\caption{(Color online) The (red) solid line and the (blue) dashed
line are constant energy contours $E_{\bf k}=\omega/2$ with the
frequency $\omega$ being $18$ meV and $55$ meV, respectively. The
dotted line is the normal state Fermi surface.
 }\label{fig8}
\end{figure}

These features of the spin excitation in LSCO can be traced further
to the evolution of the Fermi surface. The normal state Fermi
surface and the quasiparticle energy contours are plotted in Fig. 8.
The energy contour is closed and has a banana shape at low
frequencies, with the tips being just at the normal state Fermi
surface. The contribution to spin susceptibility  comes mainly from
the node-to-node excitations as the frequencies are close to zero.
Thus the IC peaks are along diagonal direction, and a clear spin gap
appears at the parallel direction. As the frequency increases, the
spin excitation along parallel direction is present. When the
frequency approaches to about 18 meV, the tip of the energy contour
reaches the hot spots (the crossing points of the Fermi surface with
the magnetic Brillouin zone boundary), as seen in Fig. 8. Thus the
bare spin susceptibility has a step-like rise at the momentum
$(\pi,\pi)$, leading to the quasi-resonance at this frequency. The
superconducting gap plays a minor role as the frequency increases
further, and the energy contour is not closed and contains two
parts: (I) and (II), as depicted in Fig. 8. The shapes of the both
parts resemble the normal state fermi surface. Part I is just like
the underdoped fermi surface, while the part II resembles the
overdoped one. As we know, the spin excitation tends to be
commensurate and the intensity increases as the doping decreases to
near the half-filled case~\cite{jxli}. Therefore, the spin
excitation from part I $\rightarrow$ I also tends to be commensurate
as the frequency increases, with the intensity being larger than the
excitation from part II $\rightarrow$ II, leading to the
commensurate spin excitation occurring at the frequency 55 meV.
While spin excitation from part II to II also contributes to the
spectra, and the peak at this frequency is broader than the lower
frequency IC peak at 18 meV.

\section{summary}

Based on the fermiology theories, we have examined the evolution of
the spin susceptibility for different systems near the optimal
doping. The spin excitation is dominated by a resonance peak in the
superconducting state as $T_c$ is high enough, with the resonance
frequency being proportional to $T_c$. The peak intensity becomes
weaker as $T_c$ is lower and only quasi-resonance occurs as $T_c$
decreases to 40 K. For LSCO samples, the presence of the spin-stripe
order suppresses the resonance further, and the IC peaks develop at
the expected resonance frequency (about 18 meV), with the intensity
being still enhanced by the renormalized effect of the RPA factor.
The intensity of the IC peaks decreases as the frequency increases
from 18 meV, then it increases again and reaches the local maximum
at the frequency 55 meV, with the spin excitation being commensurate
at this frequency. This two-frequency component is originated from
the competition between the RPA factor and the bare spin
susceptibility. The dispersion of the spin excitation behaves a
hourglass shape in both the superconducting and normal states. We
also note that the spin excitation may be one dimensional due to the
one-dimension spin-stripe order. For YBCO samples, the spin
excitation is significantly different from that of LSCO, namely, a
clear resonance peak with a large intensity appears at about 40 meV.
The higher frequency tail observed in LSCO is not seen in YBCO. On
the other hand, the dispersion in the superconducting state is
similar to that of LSCO, while it is different in the normal state,
namely, the spin excitation is commensurate in the normal state for
all the frequencies considered. The present results are consistent
with the experiments. We have elaborated all the results based on
the topology of the Fermi surface.

\begin{acknowledgments}
This work was supported by the RGC grants of Hong
Kong (HKU 7050/03P and HKU-3/05C), the NSFC (10429401), and the
973 project of China (2006CB601002).

\end{acknowledgments}

\end{document}